\documentclass[12pt]{article}

\usepackage[utf8]{inputenc}
\usepackage{cancel}
\usepackage{enumitem}
\usepackage{amssymb}
\usepackage{soul}
\usepackage{tabularx}
\usepackage{xcolor}
\usepackage{booktabs}
\usepackage{subcaption} 
\usepackage{colortbl}
\usepackage{authblk}
\DeclareUnicodeCharacter{00A0}{ }

\newcolumntype{C}{>{\centering\arraybackslash}X}
\usepackage[T1]{fontenc}
\usepackage{epsfig}
\usepackage{latexsym}
\usepackage{graphicx}
\usepackage{amsmath}
\usepackage{amsfonts}   
\usepackage{amssymb}    
\usepackage{float}
\usepackage{bm}
\usepackage{url}
\usepackage{hyperref} 
\usepackage[nodisplayskipstretch]{setspace}
\def\lsim{\raise0.3ex\hbox{$\;<$\kern-0.75em\raise-1.1ex\hbox{$\sim\;$}}}
\def\gsim{\raise0.3ex\hbox{$\;>$\kern-0.75em\raise-1.1ex\hbox{$\sim\;$}}}

\def    \beq            {\begin{equation}}
\def    \eeq            {\end{equation}}
\def    \bea           {\begin{eqnarray}}
\def    \eea           {\end{eqnarray}}

\def \mn{\mu\nu{\rm SSM}}

\def\g2{{\rm GeV}^2}

\def\order#1{\ensuremath{{\cal O}(#1)}}

\def\sw2{sin^2 \theta_w}

\def\a^tau{\alpha_{\tau}}

\def\beq{\begin{equation}}
\def\eeq{\end{equation}}
\def\beqa{\begin{eqnarray}}
\def\eeqa{\end{eqnarray}}

\newcommand{\newc}{\newcommand}
\newc\BR{BR}
\newc{\akappa}{A_{\kappa} }
\newc\deltagmtwo{\delta (g-2)_{\mu}} 
\newc\deltaamu{\Delta a_{\mu}}

\def\anti{\overline}

\def\la{\lambda}

\newc{\haa}{BR\(h_1\to a_1 a_1\)}
\newc{\abb}{BR\(a_1\to b\anti{b}\)}
\newc{\hbb}{BR\(h_1\to b\anti{b}\)}
\newc{\abund}{\Omega h^2}
\newc\bsgamma{b\rightarrow s \gamma }
\newc\bxsgamma{\overline{B}\rightarrow X_{s}\gamma}
\newc\brbsgamma{\BR(\overline{B}\rightarrow X_s\gamma)}

\allowdisplaybreaks

\usepackage{array, makecell}
\usepackage{boldline}
\usepackage[nosort]{cite}

\title{\bf{Right-handed neutrinos, domain walls and tadpoles in the superstring inspired $\mu\nu$SSM
}}

\author[a,b]{D.~E.~L\'opez-Fogliani\thanks{daniel.lopez@df.uba.ar}}
\author[c,d]{C.~Mu\~noz\thanks{c.munoz@uam.es}} 

  \affil[a]{Instituto de F\'isica de Buenos Aires UBA \& CONICET, Departamento de F\'isica, Facultad de Ciencia Exactas y Naturales, Universidad de Buenos Aires, 1428 Buenos Aires, Argentina}
  \affil[b]{Pontificia Universidad Cat\'olica Argentina, 1107 Buenos Aires, Argentina}
    \affil[c]{Departamento de F\'{\i}sica Te\'{o}rica, Universidad Aut\'{o}noma de Madrid (UAM), Campus de Cantoblanco, 28049 Madrid, Spain}
  \affil[d]{Instituto de F\'{\i}sica Te\'{o}rica (IFT) UAM-CSIC,  Campus de Cantoblanco, 28049 Madrid, Spain}

\date{}

\begin{document}

\maketitle

\begin{abstract}

We discuss the special role of right-handed neutrinos in the $\mu\nu$SSM for solving the $\mu$- and $\nu$-problems, simultaneously avoiding the potential domain wall and tadpole problems, 
In particular, embedding the $\mu\nu$SSM
in the framework of superstrings implies that not all gauge invariant terms are necessarily present in the superpotential, and thus there is more flexibility to allow only those that avoid the domain wall and tadpole problems to be present. These can be non-renormalizable terms of dimension 4 or terms of higher dimensions. In addition, non-perturbative effects can also solve both problems.
We discuss another implication of the superstring inspired $\mu\nu$SSM, since
the right-handed neutrino is expected to have extra $U(1)$ charges at high energies.
In this case, 
the cubic right-handed neutrino terms in the superpotential, helpful for generating Majorana masses and solving domain wall and tadpole problems, can arise through a variety of stringy mechanisms.

\end{abstract}  


\clearpage 

\tableofcontents 

\section{Introduction}
\label{introduction}

The addition of right-handed (RH) neutrinos $\nu_{R}$ to the spectrum of the standard model (SM) provides the light neutrinos with masses at the tree level.
In the framework of supersymmetry (SUSY), the `$\mu$ from $\nu$' Supersymmetric Standard 
Model ($\mu\nu$SSM)~\cite{LopezFogliani:2005yw} (for a recent review, see Ref.~\cite{Lopez-Fogliani:2020gzo}), was proposed 
to solve the
$\mu$- and $\nu$-problems simultaneously without
the need to introduce additional energy scales beyond the SUSY-breaking scale, which is the source of the electroweak symmetry breaking (EWSB).
To perform it, the presence in the superpotential of
trilinear couplings involving RH neutrino superfields, $\nu_{i}^c$, in addition 
to the usual quark and charged-lepton Yukawa couplings, is crucial~\cite{LopezFogliani:2005yw,Escudero:2008jg}:
\bea
\label{Eq:superpotentialmunu}
W_{\text{$\mu\nu$SSM}} &=&  
Y^e_{ij} \, H_d\,  L_i \,  e_j^c
\ +
Y^d_{ij} \,  H_d\,  Q_i \,  d_j^c 
\ -
Y^u_{ij} \,  {H_u}\, Q_i \, u_j^c
\nonumber\\
&-&
Y^\nu_{ij} \, {H_u}\,  L_i \, \nu^c_{j}
\ +
\lambda_{i} \,  {H_u} \,  H_d \,  \nu^c_{i}
\ +
\frac{1}{3}
\kappa{_{ijk}} \, \nu^c_{i}\,  \nu^c_{j}\, \nu^c_{k}
\,.
\label{superpo}
\eea
Here the summation convention is implied on repeated indexes, with 
$i,j,k=1,2,3$ the usual family indexes of the SM.\footnote{The number of RH neutrinos is in fact a free parameter. These particles are the only SM singlets and therefore their way of arising from a more fundamental theory can be different from the other particles. An analysis using this number as a free parameter can be found in Ref.~\cite{Aguilar-Saavedra:2021qbv}.
Nevertheless, we will assume for simplicity in what follows that their number is three, replicating what happens with the other SM fermions, since this assumption will not modify our results.}
Our convention for the contraction of two $SU(2)$ doublets is e.g.
$ {H}_u \,   H_d\equiv  \epsilon_{ab}  H^a_u \,  H^b_d$,
$a,b=1,2$ and $\epsilon_{ab}$ the totally antisymmetric tensor with $\epsilon_{12}=1$.

The last three terms containing RH neutrinos in the above superpotential are clearly allowed by gauge invariance, since these fields have vanishing hypercharges by construction.
In particular, the fifth term in Eq.~(\ref{superpo}) 
is an explicit $R$-parity (and lepton-number) violating term which
generates dynamically a $\mu$-parameter after EWSB, 
$\mu=\la_i v_{iR}$, with $v_{iR}$ the vacuum expectation values (VEVs) of the right sneutrinos.
This mechanism solves the so-called
$\mu$-problem~\cite{Kim:1983dt} (for a recent review, see Ref.~\cite{Bae:2019dgg}).
It is analogue to the solution of the Next-to-Minimal Supersymmetric Standard Model
(NMSSM) (for reviews, see Refs.~\cite{Maniatis:2009re,Ellwanger:2009dp}), where instead of RH neutrino superfields an extra singlet superfield $S$ is added to the spectrum with the coupling
$\lambda H_u  H_d S$.
{By the same EWSB mechanism, 
the neutrino Yukawa term 
generates 
effectively bilinear terms, $\mu_i=Y^{\nu}_{ij} v_{jR}$,
as the ones of the Bilinear $R$-Parity Violating model (BRPV) (for a review, see Ref.~\cite{Barbier:2004ez}),
as well as
Dirac masses for neutrinos, 
$m_{{\mathcal{D}_{ij}}}= Y^{\nu}_{ij} v_u$, with $v_u$ the VEV of the Higgs $H_u$.
In addition, the cubic neutrino} term
generates electroweak (EW)-scale Majorana masses,
${\mathcal M}_{ij}
= {2}\kappa_{ijk} v_{kR}$,
instrumental in
generating the correct 
neutrino masses 
and mixing 
angles~\cite{Capozzi:2017ipn,deSalas:2017kay,deSalas:2018bym,Esteban:2018azc},
i.e. in solving 
the $\nu$ problem through an EW-scale seesaw with $Y^{\nu}_{ij}\lsim 10^{-6}$~\cite{LopezFogliani:2005yw,Escudero:2008jg,Ghosh:2008yh,Bartl:2009an,Fidalgo:2009dm,Ghosh:2010zi,Liebler:2011tp}.
The presence of this cubic term
also forbids a Peccei-Quinn-like symmetry $U(1)_{PQ}$ in the Lagragian, avoiding therefore the existence of a phenomenologically unacceptable massless Nambu-Goldstone boson once EWSB takes place.
Again, this is analogue to what happens in the NMSSM where the cubic term $\kappa  S  S  S$ is present.


Since $R$-parity
and lepton number are not conserved in the $\mn$
in contrast to the NMSSM, this leads
to a completely different
phenomenology characterized by an enlarged Higgs sector which includes the left and right sneutrinos~{\cite{Escudero:2008jg,Fidalgo:2011ky,Ghosh:2014ida,Biekotter:2017xmf,Biekotter:2019gtq,Kpatcha:2019qsz}, and distinct prompt or displaced
decays of the lightest supersymmetric particle (LSP)
producing multi-leptons/jets/photons with small/moderate missing transverse energy 
from 
neutrinos.
Studies of these 
interesting signals 
at the Large Hadron Collider (LHC) 
have been carried out in the literature~\cite{Ghosh:2017yeh,Lara:2018rwv,Lara:2018zvf,Kpatcha:2019gmq,Kpatcha:2019pve,Heinemeyer:2021opc,Kpatcha:2021nap}, taking into account the intimate connection between the LSP lifetime and the size of neutrino Yukawas.
The low decay width of the LSP due to the smallness of neutrino masses is also related to the existence of possible candidates for decaying dark matter in the model.
This is the case of 
the gravitino~\cite{Choi:2009ng,GomezVargas:2011ph,Albert:2014hwa,GomezVargas:2017,Gomez-Vargas:2019mqk}, or the axino~\cite{Gomez-Vargas:2019vci}, with lifetimes greater than the age of the Universe.
It is worth mentioning concerning cosmology, that baryon asymmetry might be realized in the
$\mn$ through EW baryogenesis~\cite{Chung:2010cd}.
See also Refs.~\cite{Escudero:2008jg,Biekotter:2021rak} for vacuum structure analyses of the $\mn$.
The EW sector of the $\mn$ can also explain~\cite{Kpatcha:2019pve,Heinemeyer:2021opc}
 the longstanding
discrepancy between the experimental result for the anomalous magnetic
moment of the muon~\cite{Abi:2021gix,Albahri:2021ixb} and the SM prediction.

However, despite all these interesting theoretical and phenomenological properties, similar to the NMSSM the $\mn$ has the potential problem of the existence of a $Z_3$ discrete symmetry in its superpotential, corresponding to a multiplication of all components of all chiral superfields by a phase
$e^{2\pi i/3}$.
Discrete symmetries when spontaneously broken can
give rise to domains of different degenerate vacua separated by 
domain walls~\cite{Zeldovich:1974uw,Kibble:1976sj,Kibble:1980mv,Vilenkin:1984ib}.
The walls would dominate the energy density in the Universe producing wall driven inflation, unless they are removed well before.
Besides, domain walls can also disrupt primordial nucleosynthesis.
These cosmological domain wall problems, if present, can be solved if the degeneracy of the vacua is slightly 
broken, eventually leading to the dominance of the true 
vacuum~\cite{Zeldovich:1974uw}. 

This mechanism to solve the domain wall problem was applied in the NMSSM
assuming that the $Z_3$ discrete symmetry is explicitly broken by gravitationally supressed interactions producing
non-renormalizable terms in the superpotential, which
lift the degeneracy of the three original vacua~\cite{Ellis:1986mq,Rai:1992xw,Abel:1995wk}.\footnote{Once this solution is present, the domain walls were used in Ref.~\cite{Abel:1995uc} for the generation of baryogenesis in the NMSSM.} Besides, these terms can be chosen small enough as not to alter the low-energy 
phenomenology.
However, another problem arises when working 
in the framework of NMSSM supergravity. 
The presence of a linear term
$t S$ in the superpotential (with $t$ of dimension mass squared) would destabilized the hierarchy since the mass scale is expected to be proportional to the cut-off scale of the theory, which is typically the Planck mass $M_{\text{Pl}}\approx 1.2\times 10^{19}$ GeV.
Although its presence is forbidden by the $Z_3$ symmetry,
the non-renormalizable terms together with soft SUSY breaking can reintroduce it in the effective potential. Such corrections to the potential are quadratically divergent and therefore proportional to powers of $M_{\text{Pl}}$, destabilizing the hierarchy~\cite{Ellwanger:1983mg,Bagger:1993ji,Jain:1994tk,Bagger:1995ay,Abel:1995wk,Abel:1996cr}.

Any supergravity model with gauge singlets can have this tadpole problem, since the dangerous diagrams are not excluded by gauge invariance.
It has been argued nevertheless in the framework of the NMSSM that the non-renormalizable terms generating them are not necessarily present when the model possesses either target space duality in a string effective 
action~\cite{Abel:1996cr} or $R$-symmetry~\cite{Abel:1996cr,Panagiotakopoulos:1998yw,Panagiotakopoulos:1999ah}, but other terms that do not destabilize the hierarchy, and solve in addition the domain wall problem, can be present.

The aim of this work is to analyze domain wall and tadpole problems in the framework of the 
$\mn$, where SM singlets such as RH neutrinos are present. In particular, we will see that these problems can be solved naturally in the context of superstring compactifications, where not all gauge invariant terms are necessarily allowed. 
{Besides, non-perturbative effects can also be relevant.}
The paper is organized as follows. 
Sec.~\ref{model} will be devoted to discuss the advantage of constructing the $\mn$ in the framework of superstring theory.
In Sec.~\ref{domain}, we will analyze how the presence of non-renormalizable terms can be helpful for solving the domain wall problem.
In Sec.~\ref{tadpole}, we will see that these terms can induce simultaneously destabilizing tadpole divergences, and we will study how to solve this additional problem. 
Finally, in Sec.~\ref{neutrinos} we will explain first that RH neutrinos, whose presence is crucial in the framework of the $\mn$, can have extra $U(1)$ charges in string compactifications. Then, we will discuss how these charges modify the origin of certain terms in the superpotential with interesting consequences.
Our conclusions are left for Sec.~\ref{conclusion}.

\section{The $\mn$ and renormalizable terms}
\label{model}

In this section we will discuss how natural is the presence in $W_{\text{$\mu\nu$SSM}}$
of only the terms shown in Eq.~(\ref{Eq:superpotentialmunu}), when 
RH neutrinos are added to the spectrum of the SUSY SM.
Note that the existence of the latter fields
imply that the $SU(3)\times SU(2)\times U(1)_Y$ gauge invariant superpotential containing the most general renormalizable terms is in fact given by:
\bea
\label{Eq:superpotential10}
W_r &=&  
Y^e_{ij} \, H_d\,  L_i \,  e_j^c
\ +
Y^d_{ij} \, H_d\, Q_i \, d_j^c 
\ -
Y^u_{ij} \, {H_u}\,  Q_i \,  u_j^c
\ +
\mu \,  H_u \,  H_d
\nonumber\\
&+&
\lambda_{ijk} \, L_i  \, L_j \,  e_k^c
+
\lambda'_{ijk} \,  L_i \, Q_j \, d_k^c
\ +
\lambda''_{ijk} \,  u_i^c \,  d_j^c \,  d_k^c
\ + \mu_i \,  H_u \,  L_i
\nonumber\\
&-&
Y^\nu_{ij} \,  {H_u}\, L_i \, \nu^c_{j}
\ + \lambda_{i} \, {H_u} \, H_d \, \nu^c_{i}
\ +
\frac{1}{3}
\kappa{_{ijk}} \, \nu^c_{i}\, \nu^c_{j}\, \nu^c_{k}
\ +
{\mathcal M}_{ij} \, \nu^c_{i}\, \nu^c_{j}
\ + t_{i} \hat \nu^c_{i}
\,.
\eea
The four terms in the first line determine the superpotential of the Minimal Supersymmetric Standard Model (MSSM) (for reviews, see e.g. Refs.~\cite{Nilles:1983ge,Haber:1984rc,Martin:1997ns}).
The four terms in the second line are the conventional trilinear and bilinear
$R$-parity violating (RPV) couplings (for a review, see Ref.~\cite{Barbier:2004ez}). Finally, all the terms in the last line contain RH neutrino superfields.
The first three terms are characteristic of the $\mu\nu$SSM as discussed in the Introduction, the fourth one is a bilinear term giving Majorana masses for RH neutrinos, and the fifth one is a linear (tadpole) term for them.

{Given superpotentials (\ref{Eq:superpotentialmunu}) and (\ref{Eq:superpotential10}),
one would like to have an explanation for the absence in $W_{\text{$\mu\nu$SSM}}$ of some of the terms present in $W_r$. In particular, this is the case of the dangerous tadpole terms $t_{i} \nu_{i}^c$, as well as of the mass terms
$\mu H_u H_d$, $\mu_i H_u L_i$ and 
${\mathcal M}_{ij} \nu_{i}^c\nu_{j}^c$ which would reintroduce 
the $\mu$-problem and additional naturalness problems if they are of the order of the high-energy scale. 
Here we are interested in the analysis of the $\mn$ inspired in the construction of superstring models, 
given the relevance of string theory as a possible underlying unified theory. In this framework, not all gauge invariant terms are necessarily allowed.
In particular, at the perturbative level in string constructions the massive modes have typically huge masses of the order of the string scale, whereas the massless ones give rise only to trilinear terms at the renormalizable level.
Thus one ends up with an accidental $Z_3$ symmetry in the low-energy theory. 
Thanks to it, the presence of the above dangerous terms 
is automatically forbidden.
Let us remark nevertheless that using $W_{\text{$\mu\nu$SSM}}$ these terms arise dynamically with the correct (EW) scale after EWSB, as discussed in the Introduction.
This fact motivates the use of the superpotential (\ref{Eq:superpotentialmunu}), where only dimensionless parameters contribute.}

On the other hand, it is well known that the simultaneous presence of the terms
$\lambda'_{ijk} L_i  Q_j  d_k^c$  and 
$\lambda''_{ijk}  u_i^c  d_j^c   d_k^c$, violating lepton and baryon number respectively,
is dangerous since they would produce fast proton decay. The usual assumption in the literature of the MSSM is to invoke an {\it ad hoc} $Z_2$ discrete symmetry $R$-parity \cite{Barbier:2004ez} to avoid the problem,  providing with charge
 $+1$ to ordinary particles and $-1$ to their superpartners.
However, this assumption is 
clearly too stringent, since then other terms like e.g. $\lambda_{ijk}  L_i   L_j  e_k^c$ or $\lambda_{i}\nu_{i}^c H_u  H_d$, which are harmless for proton decay, would also be forbidden.
Less drastic solutions arise in the context of string theory. In compactifications of the heterotic string on orbifolds~\cite{Dixon:1985jw} the matter fields lie in different sectors of the compact space. For example, for the case of the $Z_3$ orbifold couplings between three fields in the untwisted sector or three fields in the twisted sector are allowed, but couplings between untwisted and twisted fields are forbidden by the space group selection rules. Similar results about couplings that are forbidden even if they are gauge invariant, can also be obtained in other kind of compactifications and in other types of strings. Thus ``stringy'' selection rules can forbid couplings that in principle seem to be allowed by gauge invariance, as pointed out in Ref.~\cite{Escudero:2008jg}.
As a consequence, the RPV couplings $\lambda''_{ijk}$ can be naturally forbidden whereas the other RPV couplings
are allowed.
In fact, the couplings $\lambda_{ijk}$ and $\lambda'_{ijk}$ can be naturally added to  
$W_{\text{$\mu\nu$SSM}}$~\cite{LopezFogliani:2005yw,Escudero:2008jg,Ghosh:2017yeh}.

Alternatively, in the context of e.g. heterotic orbifolds or free fermion constructions the typical presence of
several $U(1)$ gauge symmetries beyond the hypercharge (see a more detailed discussion below in Sec.~\ref{neutrinos}), can also be useful. Although these extra $U(1)$'s are spontaneously broken by VEVs of scalar fields through $D$- and $F$-flat directions of the scalar potential, dangerous couplings allowed by the SM gauge symmetry as those discussed 
above can be forbidden by the extra $U(1)$ charges~\cite{Casas:1987us,Casas:1988hb,Casas:1988vk}.
In fact, residual $Z_N$ symmetries can be left in the low-energy theory depending on the choice of flat direction.
In Refs.~\cite{Berasaluce-Gonzalez:2011gos,Ibanez:2012wg}, a less model-dependent mechanism was obtained in type IIA intersecting brane constructions, where e.g. models with $Z_3$ Baryon-parity which forbids 
$\lambda''_{ijk}$ couplings were constructed.


We conclude therefore that string constructions facilitate the presence of only certain couplings in $W_{\text{$\mu\nu$SSM}}$ solving crucial theoretical problems, such as the $\mu$- and $\nu$-problems.
However, this superpotential has a $Z_3$ discrete symmetry
that can induce a cosmological domain wall 
problem. This is the subject that we will discuss in the next sections. As we will see, the addition of non-renormalizable terms to $W_{\text{$\mu\nu$SSM}}$ breaks the $Z_3$ symmetry and can generate dynamically the appropriate contributions to solve this problem, simultaneously avoiding the destabilization of the hierarchy due to tadpoles.

Let us finally comment that there is
another strategy to explain the absence of linear, bilinear and baryon-number violating terms in $W_{\text{$\mu\nu$SSM}}$, solving simultaneously the domain wall {and tadpole problems,
without relying in the superstring framework.}
It
consists of extending 
the SM gauge group with an extra $U(1)$ at low energies~\cite{Fidalgo:2011tm,Lozano:2018esg,Aguilar-Saavedra:2021qbv}
(see Ref.~\cite{Langacker:2008yv} for a review in other models).\footnote{
This strategy is also possible in string constructions, where as mentioned above several $U(1)$'s are typically present and some of them could survive at low energies (see e.g. the orbifold model with an extra $U(1)$ of Ref.~\cite{Casas:1988se}).}
Then, the dangerous operators could be forbidden by gauge invariance, and the domain wall problem disappears once the discrete symmetry is embedded in the broken gauge 
symmetry~\cite{Lazarides:1982tw,Kibble:1982dd,Barr:1982bb}. 
After the EW phase transition one expects a network of domain walls bounded by cosmic strings to form and then collapse.
{Since the RH neutrinos (and the other matter fields) are charged under the extra $U(1)$ the dangerous tadpole diagrams could also be forbidden.}
In addition to the presence of the extra $U(1)$ at low energies, another
crucial characteristic of the models in Refs.~\cite{Fidalgo:2011tm,Aguilar-Saavedra:2021qbv} is the presence in their spectrum of exotic matter dictated by the anomaly cancellation conditions, such as extra quark representations or singlets under the SM gauge group.
In addition to the non-minimality of the particle spectrum,
gauge invariance forbids the cubic term 
$\kappa_{ijk} \nu_{i}^c\nu_{j}^c\nu_{k}^c$ that generates Majorana masses, making more involved to obtain the correct neutrino masses and mixing angles.

In the following we will focus our analysis in the minimal $\mn$ characterized by the SM gauge group and the superpotential
(\ref{Eq:superpotentialmunu}), in the framework of superstrings.

\section{Domain wall problem and non-renormalizable terms}
\label{domain}
 
 The $Z_3$ discrete symmetry of the superpotential of the $\mn$ is spontaneously broken due to the EWSB in the early Universe. As discussed in the Introduction, in this case dangerous domain walls separating degenerate vacua can be produced. 
 The strongest constraint on domain walls arises from nucleosynthesis, since the walls must disappear before the nucleosynthesis era beginning at $T\sim 1$ MeV.
 If we trust this cosmological argument, to solve the problem requires the presence in the effective potential of $Z_3$-breaking terms~\cite{Abel:1995wk}
\bea
\delta V \gsim 10^{-7} \
\frac{v_R^3 m^2_W}{M_{\text{Pl}}},
 \label{effpot}
\eea
where the surface energy of the domain wall $\sigma\sim v^3$, with
$v$ a typical VEV of the fields determining the scale of the EWSB,
{has been approximated using in the $\mn$ $v\sim v_{iR}\sim v_R$}.
This condition ensures that the pressure 
produced by one of the vacua slightly deeper than the others, $\epsilon\sim \delta V$, is larger than the surface tension $\sigma/R$ (where $R$ is the curvature scale of the wall) before nucleosynthesis, leading to the dominance of the true vacuum.

Now, one can straightforwardly see that non-renormalizable 
dimension-n terms, with $n>3$, in the superpotential (working with the K\"ahler potential will not modify our conclusions)
\bea
W_{\text{nr}} = c_n\ \frac{
\phi_1\cdot\cdot\cdot\phi_n
}{\Lambda^{n-3}},
\label{n.r.}
\eea
break explicitly the discrete $Z_3$ symmetry
inducing the following soft SUSY-breaking contributions to the effective potential
\bea
\delta V_{\text{nr}} \approx c_n\ m_{3/2}  \frac{
v_1
\cdot\cdot\cdot
v_n
}{\Lambda^{n-3}},
 \label{effpotnr}
\eea
where $v_{1,...,n}$ are the VEVs of the fields $\phi_{1,...,n}$, we have assumed soft SUSY-breaking parameters $\sim m_{3/2}$, and 
$\Lambda$ is the cut-off scale of the high-energy theory.
To solve the domain wall problem, $\delta V_{\text{nr}}$ must be large enough to 
fulfil requirement~(\ref{effpot}).

In the case of a supergravity theory where $\Lambda=M_{\text{Pl}}$,
obviously only dimension-4 terms producing  
\bea
\delta V_{\text{nr}} \approx c_4\ \frac{
v_R^5
}{M_{Pl}},
 \label{effpotnr23}
\eea
can fulfill this requirement
with the coupling constant verifying the reasonable bound,
$c_4\gsim10^{-9}$,
where we have used $m_{3/2}\sim v_R\sim v_{1,...,4}\sim 1$ TeV.
Terms of dimension larger than four are too much suppressed by the Planck mass.
For example, a dimension-5 term in the superpotential 
$c_5\phi^{5}/M_{\text{Pl}}^2$ gives rise to
$\delta V_{\text{nr}} \approx c_5 v_R^6/M_{Pl}^2$,
implying the
bound
$c_5\gsim 10^{7}$,
well beyond perturbativity theory.

On the other hand, even in superstrings the relevant scale of the theory can be much lower than $M_{Pl}$ (see e.g. Ref.~\cite{Ibanez:1998rf} and references therein).
As pointed out in Ref.~\cite{Chung:2010cd}, as long as $\Lambda < M_{\text{Pl}}$ 
the above result on the dimensionality of the non-renormalizable terms to solve the domain wall problem, $n=4$, 
can be relaxed.
In particular, from Eqs.~(\ref{effpot}) and~(\ref{effpotnr}) it is straightforward to deduce that in general the coupling constants must verify
\bea
c_n\gsim 10^{-25} \left(\frac{\Lambda}{v_R}\right)^{n-3}.
\label{cn}
\eea
Thus, even for values of the cut-off scale as large as $\Lambda = 10^{15}$ GeV, one obtains that dimension-5 terms can be used to solve the domain wall problem. 
For very low values of the scale such as $\Lambda = 10^4$ GeV, terms of dimension as large as 27 can be used. 
In the following we will consider only the case $\Lambda=M_{Pl}$, but the results can be straightforwardly extrapolated for lower scales.

To solve the domain wall problem in the $\mn$ supergravity model we have available the following 
dimension-4 terms in the superpotential, with the only contributions of RH neutrino and Higgs superfields: 
\bea
\label{nr4}
\kappa^{\text{nr}}_{ijkl} \,  \nu^c_{i}\,  \nu^c_{j}\,  \nu^c_{k}\,  \nu^c_{l}/M_{\text{Pl}},
\quad
\lambda^{\text{nr}}_{ij} \,  \nu^c_{i} \,  \nu^c_{j} \,  {H_u} \,  H_d/M_{\text{Pl}},
\quad
\lambda^{\text{nr}}_{ud} \, ( {H_u} \,  H_d)^2/M_{\text{Pl}},
\label{nonrenor1}
\eea
and the terms
\bea
\label{nr4n}
Y^{\nu,\text{nr}}_{ijk} \,  \nu^c_{i}\,  \nu^c_{j}\,  {H_u}\,  L_k /M_{\text{Pl}},
\quad
\kappa^{\text{nr}}_{ij}
\, {H_u}\,  L_i \, {H_u}\,  L_j/M_{\text{Pl}},
\quad
Y^{\nu,\text{nr}}_{i}
\,  {H_u}\,  L_i \,  {H_u}\, H_d/M_{\text{Pl}}.
\label{yukawas}
\eea
with the contribution also of lepton doublet superfields $L_i$.

The terms in Eq.~(\ref{nr4})
are similar to the ones available in the NMSSM supergravity model~\cite{Abel:1995wk},
substituting in the first two the RH neutrino superfields  $\nu_{i}^c$ by the singlet superfield $S$.
Now, as discussed above, to solve the domain wall problem with some of these non-renormalizable terms their couplings must fulfill requirement~(\ref{effpot}), where now
\bea
\delta V_{\text{nr}} \approx \kappa^{\text{nr}}_{ijkl}\ \frac{v_R^5}{M_{Pl}},\,\,\,\,
\lambda^{\text{nr}}_{ij}\ \frac{v_R^3 v_u v_d}{M_{Pl}},\,\,\,\,
\lambda^{\text{nr}}_{ud}\ \frac{v_R v_u^2 v_d^2}{M_{Pl}},
 \label{effpotnr3}
\eea
implying
the reasonable bounds shown at the top of Table~\ref{table:couplings}.
Obviously, the bound for the couplings $\kappa^{\text{nr}}_{ijkl}\gsim 10^{-9}$ is the same as for the $c_4$ coupling obtained in the generic discussion of Eq.~(\ref{effpotnr23}).
The stronger constraints for the other couplings,
$\lambda^{\text{nr}}_{ij}\gsim 10^{-7}$ and 
$\lambda^{\text{nr}}_{ud}\gsim 10^{-5}$,
are due to the smaller values of the 
VEVs $v_{u,d}\sim 10^{-1} v_R$.
Let us point out that these values for the couplings are reasonable, since one expects the non-renormalizable couplings to be smaller than the renormalizable ones. This is in fact the natural situation in string compactifications where the former get a suppression factor with respect to the latter (see e.g. the discussion of couplings in the $Z_3$ orbifold compactification of the heterotic string in Ref.~\cite{Casas:1989qx}).

Concerning the non-renormalizable terms in Eq.~(\ref{nr4n}), let us remark that because of RPV in the $\mn$ not only the right sneutrinos and Higgses acquire VEVs, but also the left sneutrinos in $L_i$. This is the reason for the presence of these terms in the $\mn$ unlike the case of the NMSSM.
Their contributions to $\delta V_{\text{nr}}$ depend therefore on the VEVs of the left sneutrinos which are small
$v_{iL}\sim v_L \lsim 10^{-4}$ GeV~\cite{LopezFogliani:2005yw,Escudero:2008jg,Lopez-Fogliani:2020gzo}, because their minimization equations are determined by the small neutrino Yukawas. We have respectively
\bea
\delta V_{\text{nr}} \approx Y^{\nu,\text{nr}}_{ijk}\ \frac{v_R^3 v_L v_u}{M_{Pl}},\,\,\,\,
\kappa^{\text{nr}}_{ij}\ \frac{v_R v_L^2 v_u^2}{M_{Pl}},\,\,\,\,
Y^{\nu,\text{nr}}_{i}\ \frac{v_R v_L v_u^2 v_d}{M_{Pl}},
 \label{effpotnr2}
\eea
implying 
that the bounds on the couplings are stronger, with 
$Y^{\nu,\text{nr}}_{ijk}\gsim 10^{-1}$ as shown at the top of Table~\ref{table:couplings}
and
$\kappa^{\text{nr}}_{ij}\gsim 10^{7}$,
$Y^{\nu,\text{nr}}_{i}\gsim 10$.
Because of perturbativity, clearly only the couplings $Y^{\nu,\text{nr}}_{ijk}$ 
are
useful for solving the domain wall problem.

\begin{table}[t!]
\begin{center}
\begin{tabular}{|c |c| c| c |c |}
\hline 
$\text{DW}$
&
$\kappa^{\text{nr}}_{ijkl}\gsim 10^{-9}$
& $\lambda^{\text{nr}}_{ij}\gsim 10^{-7}$ &  
$\lambda^{\text{nr}}_{ud}\gsim 10^{-5}$
& $Y^{\nu,\text{nr}}_{ijk}\gsim 10^{-1}$  
\\
$\text{T}$
&
$\kappa_{ijk}\lsim 10^{-3}$& 
$\lambda_{i}\lsim 10^{-5}$
& 
$\kappa_{ijk} \lambda_j\lambda_k\lsim 10^{-5}$
& $Y^{\nu}_{ij}\lsim 10^{-11}$ \\
\hline
\end{tabular}
\end{center}
\caption{(top) Bounds on some of the couplings to solve the domain wall (DW) problem with the corresponding non-renormalizable terms of Eqs.~(\ref{nonrenor1}) and~(\ref{yukawas}), as discussed in Sec.~\ref{domain}. (bottom) Bounds on the associated renormalizable couplings of Eq.~(\ref{superpo}) to solve simultaneously the tadpole (T) problem, as discussed in Sec.~\ref{sec4.1}.}
\label{table:couplings}
\end{table}

Thus, we have shown above that the domain wall problem 
can be solved in the framework of supergravity by dimension-4 terms. However, the latter
induce in the effective potential linear terms that generate
quadratic tadpole divergences, which destabilize the hierarchy.
This will be the subject of the next section, where we will also explore another solution to the domain wall problem through higher dimensional terms harmless to the gauge hierarchy.

\section{Solving the tadpole problem}
\label{tadpole}

 \subsection{Non-renormalizable dimension-4 terms}
 \label{sec4.1}

Given that we are working with a non-renormalizable effective theory valid below the cut-off scale $M_{\text{Pl}}$, the presence of quadratically divergent tadpole diagrams giving rise to contributions to the effective potential proportional to $M_{\text{Pl}}$ can destabilize the hierarchy~\cite{Ellwanger:1983mg,Bagger:1993ji,Jain:1994tk,Bagger:1995ay,Abel:1995wk,Abel:1996cr}. In particular, 
the first and second terms in Eq.~(\ref{nr4}) give rise to
quadratically divergences at 2-loop order
when SUSY is spontaneouly broken, as shown in Fig.~\ref{two-loop-diagram}.
In the case of the third term the tadpole divergence arises at 3-loop order.
Similar comments apply to the terms in Eq.~(\ref{nr4n}). In particular, the first one which is useful for solving the domain wall problem generates a tadpole divergence at 2-loop order
(exchanging in Fig.~\ref{two-loop-diagram} two propagators $\nu^c$ by $H_u$ and $L$).

To simplify the notation, let us denote generically all the couplings of the 2-loop divergent terms, $\kappa^{\text{nr}}_{ijkl}, \lambda^{\text{nr}}_{ij}$ and $Y^{\nu,\text{nr}}_{ijk}$ as $c_4$, and the related renormalizable couplings  contributing to the dangerous Feyman diagrams,
$\kappa_{ijk}$, $\lambda_i$ and $Y^{\nu}_{ij}$, respectively,
as $c_3$. 
Similarly, for the 3-loop divergent term we denote $\lambda^{\text{nr}}_{ud}$ as $c_4$, but there are now three renormalizable couplings
$c_3$ contributing to the Feynman diagram, two contributions from $\lambda_i$ and one from $\kappa_{ijk}$.

Following the analyses of 
Refs.~\cite{Ellwanger:1983mg,Bagger:1993ji,Jain:1994tk,Bagger:1995ay,Abel:1995wk,Abel:1996cr},
the quadratically divergent 2-loop integral of
${\order{M^2_{\text{Pl}}/(16\pi^2)^2}}$
generates in the effective potential a linear term of the form
\bea
\label{tadpoles}
\delta V\sim \frac{c_4 c_3}{(16\pi^2)^2}m_{3/2}^2 M_{\text{Pl}} (\widetilde\nu_R + \widetilde\nu^*_R).
\label{tadpolec}
\eea
For the 3-loop divergences, this result is valid but with the substitution  $c_3/(16\pi^2)^2\rightarrow c_3^3/(16\pi^2)^3$.
Let us remark related to the discussion of Eq.~(\ref{cn}) that
we are using $M_{\text{Pl}}$ as the cut-off scale of our theory.
Otherwise, we should substitute in Eq.~(\ref{tadpolec}) $M_{Pl}\rightarrow \Lambda^2/M_{Pl}$.

 \begin{figure}
    \centering
      \includegraphics[height=5.8cm]{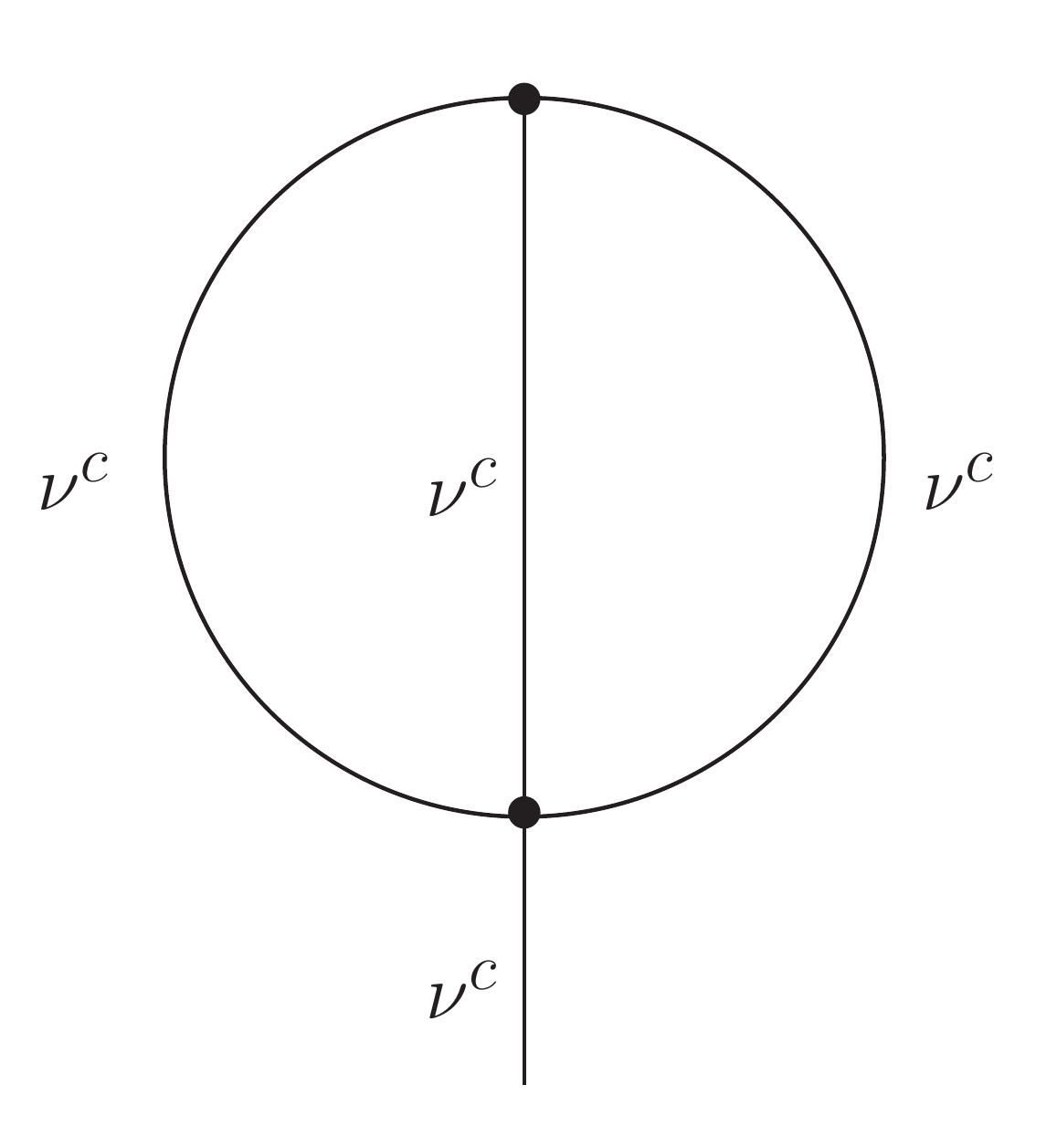}
    \caption{The dangerous diagram for the first term in Eq.~(\ref{nr4}) which can destabilize the hierarchy. For the second term the figure is the same exchanging two propagators $\nu^c$ by $H_u$ and $H_d$.}
    \label{two-loop-diagram}
    \end{figure}

Now, it is straightforward to see that the contribution (\ref{tadpolec}) destabilizes the hierarchy unless the factor in front of the RH sneutrino field is 
\bea
\label{tadpoles}
\frac{c_4 c_3}{(16\pi^2)^2}m_{3/2}^2 M_{\text{Pl}} \lsim m_{3/2}^3,
\label{tadpolec2}
\eea
implying the following bound to avoid the tadpole problem:
\bea
c_4c_3\lsim 10^{-12}. 
\eea
This constraint is compatible with requirements for the non-renormalizable couplings $c_4$ in Table~\ref{table:couplings},
$\kappa^{\text{nr}}_{ijkl}$, 
$\lambda^{\text{nr}}_{ij}$ and 
$Y^{\nu,\text{nr}}_{ijk}$,
if the related renormalizable couplings $c_3$, i.e. $\kappa_{ijk}$,
$\lambda_{i}$ and $Y^{\nu}_{ij}$ respectively,
fulfill the bounds shown at the bottom of the table.
Note, however, that the bound $Y^{\nu}_{ij}\lsim 10^{-11}$ is very unlikely to be compatible with $Y^{\nu,\text{nr}}_{ijk}\gsim 10^{-1}$,
because one expects the renormalizable couplings to be larger than the non-renormalizable ones, as discussed below Eq.~(\ref{effpotnr3}).

Concerning the other two couplings, $\kappa_{ijk}$ and $\lambda_{i}$, in the case of the NMSSM their bounds correspond to
$\kappa\lsim 10^{-3}$ and/or
$\lambda\lsim 10^{-5}$. The former would imply a light singlino, which is tightly constrained by the invisible Higgs decay and dark matter searches.
The latter would give rise to $\mu=\lambda \langle S\rangle\sim 10^{-2}$ GeV, which is excluded by experimental bounds on chargino masses.
In the $\mn$ 
there is nevertheless more flexibility since we have available 3 couplings $\lambda_{i}$ and 10 couplings $\kappa_{ijk}$. Thus hierarchies among couplings are allowed, still solving the $\mu$- and $\nu$-problems. 
For example, if $\lambda^{\text{nr}}_{1j}\gsim 10^{-7}$
is used to solve the domain wall problem, with the corresponding renormalizable coupling $\lambda_1\lsim 10^{-5}$ to avoid the tadpole problem, we can still have
$\lambda_{2,3}\lsim 1$ available for solving the $\mu$ problem.
If instead we solve the domain wall problem e.g. with $\kappa^{\text{nr}}_{112l}\gsim 10^{-9}$ 
with the corresponding $\kappa_{112}\lsim 10^{-3}$, we can have the other couplings $\kappa_{123}\lsim 1$,
$\kappa_{113}\lsim 1$, etc.,
which are useful to solve the
$\nu$-problem with the EW-scale seesaw of the $\mn$.

One might be worried that
the presence of the non-renormalizable terms 
related to these large couplings of order one, e.g. 
$\lambda^{\text{nr}}_{2j}$ or $\kappa^{\text{nr}}_{123l}$, 
would reintroduce the tadpole problem via diagrams as in 
Fig.~\ref{two-loop-diagram}.
As already discussed in Sec.~\ref{model}  not all gauge invariant renormalizable couplings have to be allowed in a particular model due to the stringy selection rules.
Similarly, not all gauge invariant non-renomalizable couplings have to be allowed. For example, three untwisted fields in the $Z_3$ orbifold of the heterotic string can only couple with a particular number of twisted fields, and three twisted fields can only couple with a particular number of twisted or moduli fields (see e.g. Ref.~\cite{Casas:1989qx} and references therein). For other $Z_n$ orbifolds the situation is similar.
These peculiarities of the superpotential also carry over to the K\"ahler potential.
For example, 
a non-minimal non-renormalizable K\"ahler term $\Phi H_uH_d/M_{\text{Pl}} + \text{hc.}$ mixing the hidden ($\Phi$) and visible ($H_{u,d}$) sectors and generating after SUSY breaking a Giudice-Masiero $\mu$-term~\cite{Giudice:1988yz}, is not present in prime orbifolds such as the $Z_3$ orbifold~\cite{LopesCardoso:1994is,Antoniadis:1995ct}. 
Thus, embedding the $\mn$ in the framework of superstrings, the stringy selections rules can forbid 
the dangerous non-renormalizable terms related to the above large couplings.


We conclude that dimension-4 terms in Eq.~(\ref{nonrenor1}) with couplings
$\kappa^{\text{nr}}_{ijkl}\gsim 10^{-9}$ solve the domain wall problem, simultaneously avoiding the tadpole problem if the corresponding renormalizable couplings fulfill 
$\kappa_{ijk}\lsim 10^{-3}$. Similarly, couplings
$\lambda^{\text{nr}}_{ij}\gsim 10^{-7}$ with the corresponding 
$\lambda_{i}\lsim 10^{-5}$ can also be used for this task.



On the other hand, for the case of 3-loop divergences the constraint becomes less stringent
\bea
c_4c^3_3\lsim 10^{-10}, 
\eea
where now
$c_4$ corresponds to $\lambda^{\text{nr}}_{ud}$
with the corresponding bound shown in Table~\ref{table:couplings},
$\lambda^{\text{nr}}_{ud}\gsim 10^{-5}$,
and $c_3^3$ to the product
$\kappa_{ijk} \lambda_j\lambda_k$.
Therefore, this constraint implies the bound $\kappa_{ijk} \lambda_j \lambda_k\lsim 10^{-5}$, as shown also in Table~\ref{table:couplings}.
For example, assuming $\kappa_{ijk}\sim \lambda_{j}\sim \lambda_k$, this constraint would imply
$\kappa_{ijk}, \lambda_{j}, \lambda_k \lsim 10^{-2}$.
Similarly to above, one could argue that depending on the concrete values of these renormalizable couplings,
the other terms generating 2-loop divergences could still destabilize the hierarchy.
But again, embedding the $\mn$ in the framework of superstring, 
the non-renormalizable terms that generate 3-loop divergences could be allowed whereas the terms generating 2-loop divergences could be forbidden.

Finally, it is worth noting that the values of the couplings discussed above are compatible with the physical vacuum, including a SM-like Higgs which in turn is compatible with collider constraints~\cite{Escudero:2008jg, Kpatcha:2019qsz,Biekotter:2021rak}. In Refs.~\cite{Escudero:2008jg, Kpatcha:2019qsz}, 
it was shown that working with small and moderate values of $\lambda_i$ and diagonal couplings $\kappa_{iii}$, i.e. $\lambda_i, \kappa_{iii}\lsim 10^{-1}$, it is 
possible to find global electroweak vacua in significant regions of the parameter space of the $\mn$.
In Ref.~\cite{Biekotter:2021rak}, the alignment-without-decoupling limit of the $\mn$ was analyzed, which implies that at least one of the $\lambda_i$ has a value of order one, close to be non-perturbative up to the GUT scale. 
In particular they used $\lambda_1\sim 0.5$, $\lambda_2\sim 0.4$ and
$\lambda_3\sim 10^{-3}$, showing that in this region of the parameter space, with all diagonal couplings $\kappa_{iii}\sim 0.4$, the possibility of long-life unstable electroweak vacua becomes viable. 
In all these results above,
negligible off-diagonal couplings $\kappa_{ijk}$ were used.
Thus, the use in
the first and third columns of Table~\ref{table:couplings} of
off-diagonal couplings $\kappa_{ijk}$
is trivially compatible with the physical minima studied
in Refs.~\cite{Escudero:2008jg, Kpatcha:2019qsz,Biekotter:2021rak}.
For example, in order to apply the solution of the third column to the vacua of 
Ref.~\cite{Biekotter:2021rak}, one can use $\lambda_1\sim 0.5$, $\lambda_2\sim 0.4$, and
$\kappa_{i12}\lsim 10^{-4}$. Diagonal couplings are also compatible, for example one could use $\kappa_{333}\sim 0.4$ with $\lambda_3\sim 10^{-3}$.
On the other hand,
the solution of the second column of Table~\ref{table:couplings} can be present in the vacua of Refs.~\cite{Escudero:2008jg, Kpatcha:2019qsz}. Also the solution of the first column with diagonal $\kappa_{iii}$.

\subsection{Non-renormalizable terms of higher dimensions}
\label{sec4.2}

We will use here the above simple argument about the flexibility of string constructions concerning the presence of certain couplings, in the context of another solution proposed to solve simultaneously the destabilization and domain wall problems. In particular, it was shown in the framework of the NMSSM~\cite{Abel:1996cr} that 
any extra odd-dimension terms in the superpotential (or even-dimension terms in the K\"ahler potential) are not harmful to the gauge hierarchy, but can avoid the domain wall problem. In the framework of the $\mn$ these are the following dimension-5 terms with the contribution of RH neutrino superfields:
\bea
\label{nr5}
c_5 \,  (\nu^c)^5/M^2_{\text{Pl}},
\quad
c_5\, (\nu^c)^3 \, ( {H_u} \,  H_d)/M^2_{\text{Pl}},
\quad
c_5 \,\nu^c \,  ( {H_u} \,  H_d)^2/M^2_{\text{Pl}},
\eea
and the terms
\bea
\label{nr5n}
c_5 \, 
(\nu^c)^3 \, ( {H_u} \,  L)/M^2_{\text{Pl}},
\quad
c_5 \,\nu^c \,  ( {H_u} \,  L)^2/M^2_{\text{Pl}},
\quad
c_5 \,\nu^c
\, ( {H_u}\, L) \, ( {H_u}\,  H_d)/M^2_{\text{Pl}},
\eea
 with the contribution of lepton doublet superfields. In these formulas we have simplified again the notation, denoting $ \nu^c\equiv\nu^c_i$, $ L\equiv L_i$, and by $c_5$ all non-renomalizable couplings.
These terms give rise to a divergence at 3-loop order
when SUSY is spontaneouly broken, as shown in Fig.~\ref{loop-diagrams}a.

 \begin{figure}
    \centering
      \includegraphics[height=4.5cm]{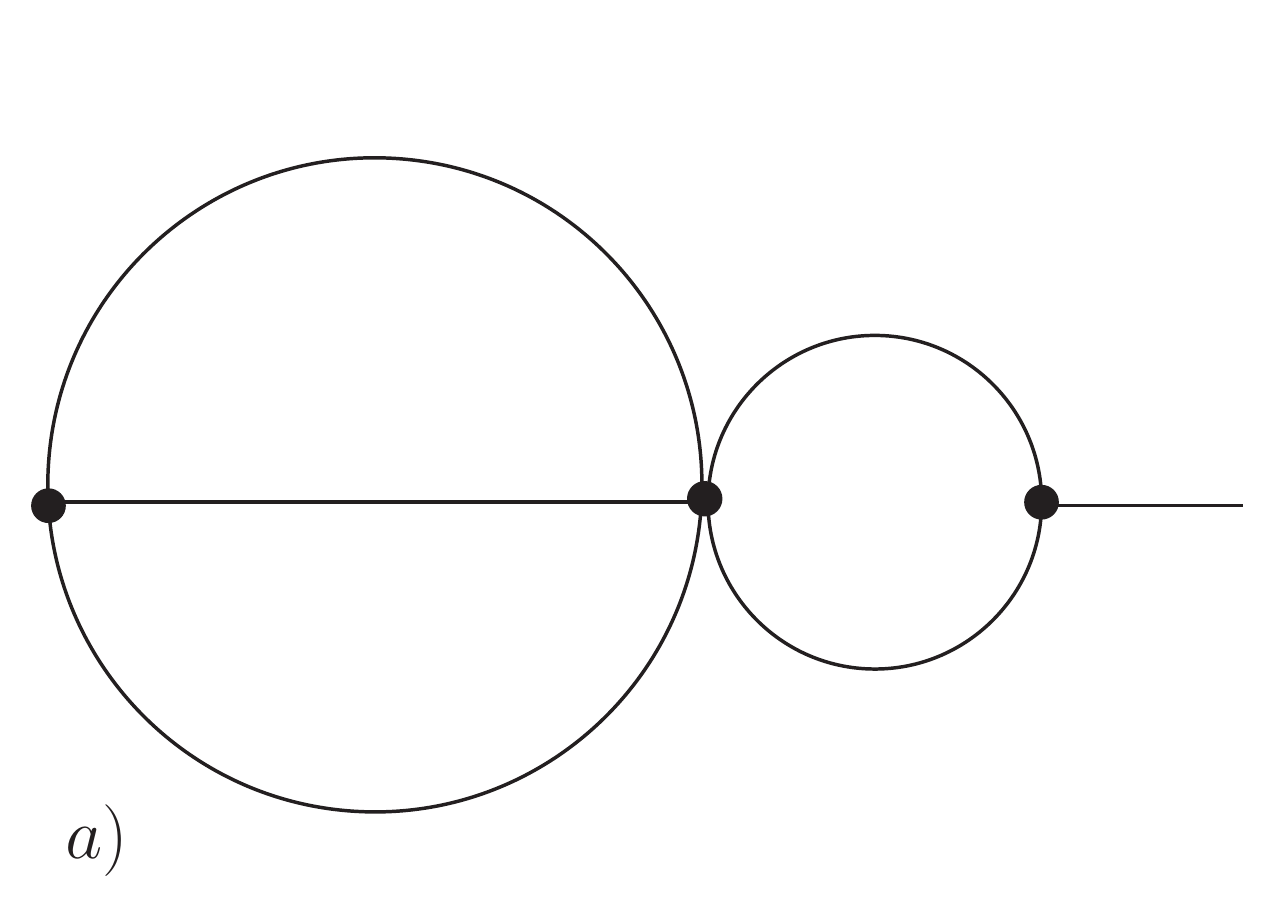}\quad \quad
       \includegraphics[height=4.5cm]{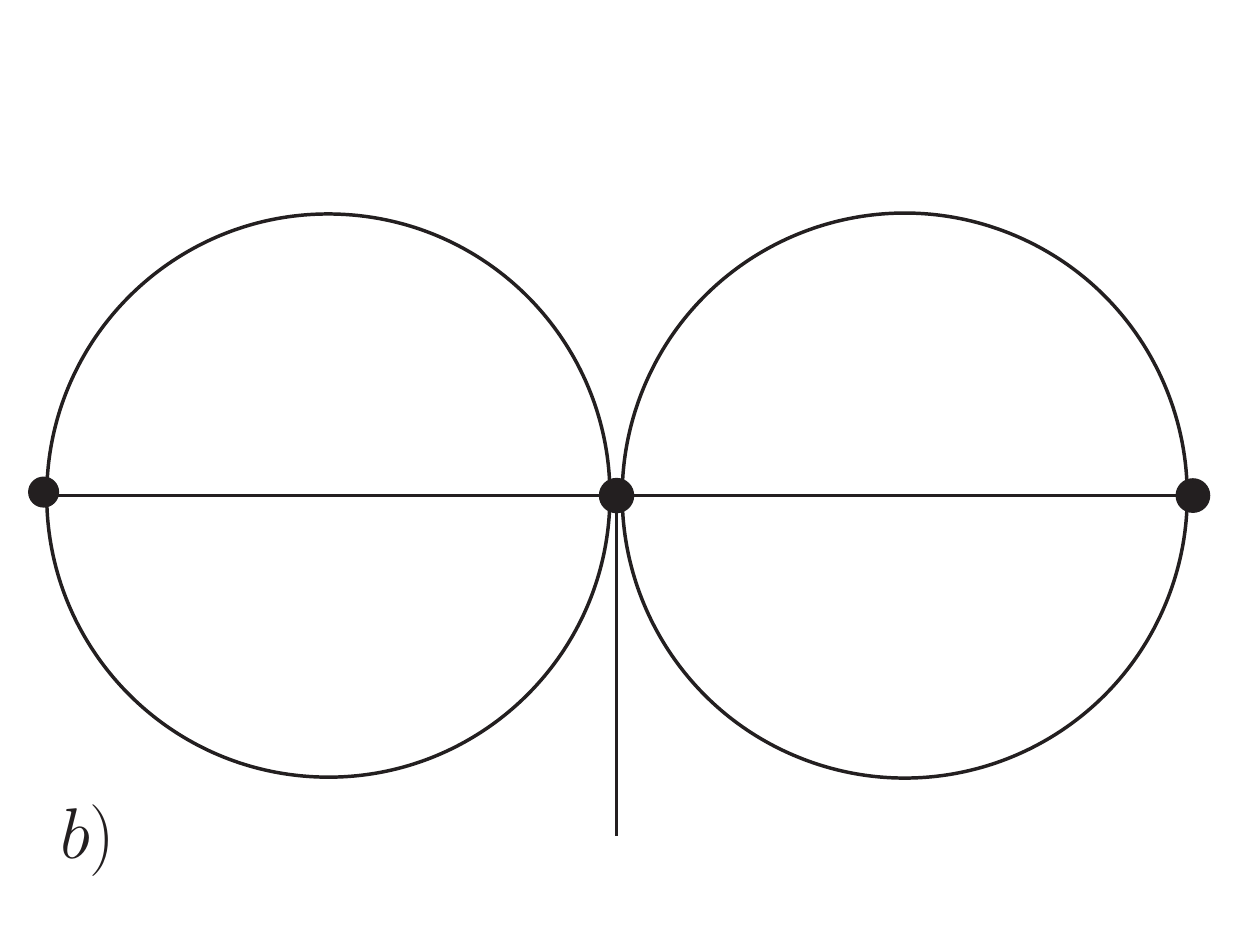}
    \caption{Tadpole diagrams for (a) dimension-5 terms in the superpotential, and (b) dimension-7 terms in the superpotential.}
    \label{loop-diagrams}
    \end{figure}

In the case of dimension-7 terms, these are
\bea
\label{nr7}
 c_7 \,  (\nu^c)^7/M^4_{\text{Pl}},
\quad
c_7\, (\nu^c)^5 \, ({H_u} \,  H_d)/M^4_{\text{Pl}},
\quad
c_7 \, (\nu^c)^3 \,  ( {H_u} \,  H_d)^2/M^4_{\text{Pl}},
\quad
 c_7 \, \nu^c \,  ( {H_u} \,  H_d)^3/M^4_{\text{Pl}},   
\eea
and
\bea
\label{nr7n}
c_7 \, 
(\nu^c)^5 \, ( {H_u} \,  L)/M^4_{\text{Pl}},
\quad
c_7 \, (\nu^c)^3 \,  ( {H_u} \, L)^2/M^4_{\text{Pl}},
\quad
c_7 \,\nu^c
\, ( {H_u}\,  L)^3/M^4_{\text{Pl}},
\nonumber
\eea
\bea
c_7 \,\nu^c
\, ( {H_u}\, L)^2 \, ( {H_u}\,  H_d)/M^4_{\text{Pl}},
\quad
 c_7 \,\nu^c
\, ( {H_u}\,  L) \, ( {H_u}\,  H_d)^2/M^4_{\text{Pl}}, &
\eea

\noindent giving rise to a divergence at 4-loop order as shown in 
Fig.~\ref{loop-diagrams}b, except the fourth term in 
Eq.~(\ref{nr7}), and the last three terms in Eq.~(\ref{nr7n}), which give rise to 
divergences at 5-loop order. 

For example, for the dimension-7 terms the quartically divergent 4-loop integral of
${\order{M^4_{\text{Pl}}/(16\pi^2)^4}}$
generates in the effective potential a linear term of the form
\bea
\label{tadpoles3}
\delta V\sim \frac{c_7 c^2_3}{(16\pi^2)^4}m_{3/2}^3 
(\widetilde\nu_R + \widetilde\nu^*_R),
\eea
which is obviously not harmful to the gauge hierarchy, but large enough to break the 
$Z_3$ symmetry (see Eq.~(\ref{effpot})), and therefore to eliminate the domain wall problem. 
Similar results are obtain for other odd-dimension terms, but with the corresponding $n$-loop factors $(16\pi^2)^{-n}$ in Eq.~(\ref{tadpoles3}).

{In the framework of the NMSSM supergravity, it was further argued that the dangerous even (odd)-dimension terms could be forbidden in the superpotential (K\"ahler potential) under special conditions.
In particular, in Ref.~\cite{Abel:1996cr} it was shown that this is possible 
when the models possess either target space duality in a string effective 
action
or gauged $R$-symmetry, and they advocated finally for generating
$\mu$-terms from couplings in the K\"ahler potential leading to small terms, 
$B\mu H_u H_d$ and $B'\mu' SS$, in the low-energy theory. 
In Ref.~\cite{Panagiotakopoulos:1998yw}, a $Z_2$ $R$-symmetry was used, allowing only the dimension-7 term $c_7  S^7/M^4_{\text{Pl}}$.
{A similar approach was used in Ref.~\cite{Ouahid:2018gpg} to attack the domain wall problem in the NMSSM extended by an $A_4\times Z_3$ flavor symmetry and three RH neutrinos}.
On the other hand, in Ref.~\cite{Panagiotakopoulos:1999ah} the discrete $Z_5$ $R$-symmetry used gave rise to the proposal of the new Minimally-extended Supersymmetric Standard Model (nMSSM)~\cite{Dedes:2000jp}. In this case, the cubic term $\kappa  S  S  S$ is forbidden by the symmetry, and its role contributing to generate the VEV of the scalar component $S$ and the breaking of the
Peccei-Quinn-like symmetry is played by the tadpole. The latter arises at 6-loop level in the effective potential by combining two non-renormalizable dimension-4 terms of the K\"ahler potential with the renormalizable superpotential term $c_3 S H_u H_d$, and is of order 
${c_4^2 c^4_3} m_{3/2}^2  M_{\text{Pl}}
(S + S^*)/(16\pi^2)^6$. 
The authors adjust ${c_4^2 c^4_3}\sim 10^{-3}$ for the tadpole to have the desired SUSY scale.
A similar model using discrete $Z_5$ and $Z_7$ $R$-symmetries, but with the tadpole contribution appearing in the superpotential, was proposed in Ref.~\cite{Panagiotakopoulos:2000wp} with the name of
Minimal Non-minimal Supersymmetric Standard Model (MNSSM).}{\footnote{It is worth noting however that global symmetries are most likely broken by quantum gravity effects~\cite{Gilbert:1989nq,Krauss:1988zc} (for a review, see e.g. Ref.~\cite{Banks:2010zn} and references therein). This is also one of the swampland conjectures (for a recent discussion, see e.g. Ref.~\cite{vanBeest:2021lhn} and references therein).}

In the case of the $\mn$ supergravity from superstrings, one has in addition the freedom of playing with the presence or absence of particular types of non-renormalizable terms.
This is model dependent, and therefore one can assume a model where only harmless odd (even)-dimension terms are allowed in the superpotential (K\"ahler potential) by the stringy selection rules, generating contributions in the effective potential of the type (\ref{tadpoles3}). These contributions are small enough as not to alter the minimization equations for the 
right sneutrinos, and therefore one can continue working with the conventional superpotential of the $\mn$ in Eq.~(\ref{Eq:superpotentialmunu}), i.e. their impact on the phenomenology can be neglected. 

If the non-renormalizable terms allowed by the stringy selection rules are of the type proposed in Ref.~\cite{Panagiotakopoulos:1999ah}, then the tadpole contribution in the $\mn$ is of the form
\bea
\label{tadpoles4}
\delta V\sim \frac{c_4^2 c^4_3}{(16\pi^2)^6}m_{3/2}^2 M_{\text{Pl}} 
(\widetilde\nu_R + \widetilde\nu^*_R),
\eea
where we have exchanged $ S \rightarrow \nu^c$.
In this case one expects ${c_4^2 c^4_3}<< 10^{-3}$, being this factor suppressed by a non-renormalizable squared coupling ($c^2_4$), and therefore this tadpole contribution solves the domain wall problem, but again without modifying the $\mn$ phenomenology. 
Even if one manages to have in a concrete model ${c_4^2 c^4_3}\sim 10^{-3}$, this contribution will not essentially modify our numerical results either since we are assuming that the stringy selection rules allow the presence of the cubic term
$\kappa{_{ijk}}   \nu^c_{i}  \nu^c_{j}  \nu^c_{k}$ in the superpotential (\ref{Eq:superpotentialmunu}).
{If this is not the case and it is not present, then to account for neutrino data becomes more involved.} 

\subsection{Non-perturbative terms}
\label{nonper}

In type II superstring another mechanism to solver the domain wall and tadpole problems is available.
As discussed in Refs.~\cite{Blumenhagen:2009qh,Cvetic:2010dz}, the singlet of the NMSSM
carries global $U(1)$ charge under the branes, and therefore perturbative couplings with several singlets are forbidden. Nevertheless, a D-instanton can generate in the superpotential non-perturbatively terms of the type 
\bea
W_{\text{np}}\sim e^{-S^{\text{cl}}} M_s^{3-n} S^n\equiv c_n S^n, 
\label{cubicc}
\eea
where
$M_s$ is the string scale, and the instanton suppression factor given by the classical instanton effective action $S^{\text{cl}}$ depends on the volumen of the three-cycle which the instanton wraps.
In Refs.~\cite{Blumenhagen:2006xt,Ibanez:2006da,Cvetic:2007ku,Ibanez:2007rs,Cvetic:2008hi},
similar instanton effects were used to generate a Majorana mass term for RH neutrinos with $n=2$.

Then, in the case of the $\mn$ a term with $n=1$ and $S\rightarrow \nu^c$ 
\bea
W_{\text{np}}\sim 
e^{-S^{\text{cl}}} M_s^{2} \nu^c, 
\label{cubicc2}
\eea
can solve the domain wall problem without spoiling the hierarchy
if the exponential suppression factor is such that $e^{-S^{\text{cl}}} M_s^2\lsim m_{3/2}^2$.

\section{Generating the cubic coupling $(\nu^c)^3$}
\label{neutrinos}

We have discussed in the previous sections the importance of the presence of RH neutrinos in the low-energy spectrum of the $\mn$ for solving the $\mu$- and $\nu$-problems, simultaneously avoiding the potential domain wall and tadpole problems, 
We would like to discuss in this section another characteristic of RH neutrinos. When added to the SM spectrum they are the only fields with no quantum numbers under the gauge group,
however in the heterotic string any SM singlet arises from 
the $E_8\times E_8$ or
$SO(32)$ gauge groups and as a consequence must be charged under some of the groups. This implies that terms of the type $(\nu^c)^3$ in the superpotential (\ref{Eq:superpotentialmunu}) are expected to be forbidden by gauge invariance. See also the discussion in Sec.~\ref{nonper} for the type II superstring, where they are also forbidden.
On the other hand, as discussed in the Introduction, this type of terms is very useful for solving the $\nu$-problem in a natural way through the generation of Majorana masses. 
We also saw in Secs.~\ref{domain} and~\ref{tadpole} that their presence can be helpful for solving the domain wall and tadpole problems.
Let us then analyze how this type of terms can originate in the superstring framework. In Ref.~\cite{Lebedev:2009ag},
the case of the singlet in the NMSSM was analyzed.

In orbifold compactifications of the heterotic string, models with $SU(3)\times SU(2)\times U(1)^n\times [\text{hidden sector}]$ as gauge group after the breaking due to the presence of Wilson lines, were built~\cite{Ibanez:1987sn}.
The matter content of these models is initially very large, including extra representations in addition to SM ones, with typically many singlets under the SM gauge group.
All these representations have non-vanishing $U(1)^n$ charges, including the RH neutrinos if present.

Subsequently, the Fayet-Iliopoulos (FI) $D$-term of the ``anomalous'' $U(1)^{(a)}$~\cite{Witten:1984dg,Dine:1987xk,Atick:1987gy,Dine:1987gj} arising from a combination of the $U(1)$'s of the models was used 
for breaking the gauge group further~\cite{Casas:1987us}. 
In particular, the anomaly is cancelled due to the Green-Schwarz mechanism~\cite{Green:1984sg}, which determines the following value of the FI $D$-term: 
\bea
D^{(a)}=c^{(a)} + \sum_\alpha Q_\alpha^{(a)} \eta_\alpha^*\eta_\alpha, \quad c^{(a)}=\frac{\sum_\alpha Q_\alpha^{(a)}}{192\pi^2} {g^{(a)}}^2 M^2_P,
\label{fidt}
\eea
where $\eta_\alpha$ are the scalar fields with charges $Q_\alpha^{(a)}$ under the anomalous 
$U(1)^{(a)}$ and $M_P\equiv M_{\text{Pl}}/\sqrt{8\pi}$.
Therefore, in order to have a SUSY vacuum solution it is necessary to give VEVs to particular scalar fields, say $\chi_\beta$, in such a way that the anomalous $D$-term is cancelled:
\bea
\sum_\beta Q_\beta^{(a)} |\langle\chi_\beta\rangle|^2 = - c^{(a)}.
\label{vevsd}
\eea
If these fields are charged under the $U(1)^n$, except under the $U(1)_Y$ which arises from a combination of the non-anomalous $U(1)$'s, then we are left at low energies with the SM gauge group in the observable sector as desired.
In addition, many of the extra matter representations acquire a high mass thus disappearing from the low-energy spectrum, giving rise to an observable sector with the usual three families of matter (plus a few extra representations)~\cite{Casas:1987us,Casas:1988hb,Casas:1988vk,Font:1988mm}.
These extra fields are related to solutions for the generation of
terms of the type $(\hat\nu^c)^3$, as we will discuss below.

\subsection{Mixing among fields}


The extra matter representations discussed above
acquire their masses 
from terms of the type:
\bea
\label{heavymass}
(\chi...\chi)\zeta, \quad
(\chi...\chi)\zeta\zeta,
\eea
where 
$\chi$ means any field acquiring a large VEV in order to cancel the FI $D$-term, and 
$\zeta$ is any field. (Of course higher-order operators must also be taken into account here.) The calculation of the massless spectrum
shows that some of the massless fields are combinations of several states~\cite{Casas:1987us,
Casas:1988hb,Casas:1988vk}.
As a consequence, some of the ``old'' physical particles have combined with other ones,
and the ``new'' particles may have more couplings allowed. 
This mechanism, relying on the mixing between fields due to FI breaking was used in Ref.~\cite{Abel:2002ih} for generating the observed structure of quark and lepton masses and mixing angles. In the case that we are interested here, 
for example $\nu^c\rightarrow \nu'^c = (a\nu^c + b\varphi^c + c\psi^c)/\sqrt{a^2+b^2+c^2}$, where $a,b,c$ are proportional to some VEVs, and $\varphi$ and $\psi$ are other singlets under $SU(3)\times SU(2)$.
Therefore, a term of the type $(\nu'^c)^3$
becomes present in the superpotential if the term $\nu^c \varphi^c\psi^c$ is initially allowed.

\subsection{Non-renormalizable couplings}

A second mechanism to generate $(\nu^c)^3$ in the superpotential is the following.
An effective coupling of massless fields which is not $U(1)^n$ invariant, say $\zeta_1\zeta_2\zeta_3$, can be generated by a higher-order operator 
\bea
\label{heavymass2}
(\chi...\chi)\zeta_1\zeta_2\zeta_3,
\eea
where as in (\ref{heavymass}) $\chi$ means any field acquiring a VEV. (Of course one has to consider here the ``new'' massless fields again.) One would expect this type of couplings that can generate $(\nu^c)^3$ as desired, to be very small being non-renormalizable and therefore suppressed by the Planck mass, but actually this is not necessarily what happens. 
Given that typically $\sum_\alpha Q_\alpha^{(a)}\gsim 100$ in the models built, one obtains from Eqs.~(\ref{fidt}) and~(\ref{vevsd}) that $\langle\chi\rangle/M_{\text{Pl}}\lsim 1$ and as a consequence effective terms $(\nu^c)^3$
of the desired order of magnitude might be generated in the superpotential of the $\mn$.
Of course, another consequence of these mechanisms it that the non-renormalizable terms discussed in the previous sections to solve the domain wall problem must be effectively generated in a similar way. 

Related to the above, let us comment the following. All the other terms in the superpotential of the $\mn$ (\ref{Eq:superpotentialmunu}) can be allowed in principle at the renormalizable level. In that case, the simultaneous presence of the terms
$Y^\nu_{ij} \, {H_u}\,  L_i \, \nu^c_{j}$ and
$\lambda_{i} \, {H_u} \,  H_d \,  \nu^c_{i}$ implies that $ H_d$ and
$ L_i$ have the same $SU(3)\times SU(2)\times U(1)^n$ quantum numbers. This seems not to be easy to obtain in the model building since one needs not only two different fields with exactly the same quantum numbers but also having different values for their couplings. We can avoid this situation if any of the above two terms arises 
from a non-renormalizable operator which is $U(1)^n$ invariant. 
In addition, this mechanism also implies that the conventional lepton-number-violating terms
$\lambda_{ijk} \,  L_i  \,  L_j \,  e_k^c$ and
$\lambda'_{ijk} \,  L_i \,  Q_j \,  d_k^c$
are forbidden (at least at the renormalizable level) once we impose the presence of the Yukawa couplings
$Y^e_{ij} \, H_d\, L_i \,  e_j^c$
and 
$Y^d_{ij} \,  H_d\,  Q_i \,  d_j^c$ in the superpotential.

\subsection{Instanton effects}

As discussed in Sec.~\ref{nonper}, instanton effects in the type II superstring can generate terms coupling NMSSM singlets among themselves (see Eq.~(\ref{cubicc}))~\cite{Blumenhagen:2009qh,Cvetic:2010dz}.
In Ref.~\cite{Aparicio:2012vk}, similar instanton effects were used to generate the cubic coupling of the NMSSM in F-theory GUTs.
This mechanism, which lies in 
the violation of the (typically anomalous) $U(1)$'s under which $S$ is charged by the instanton corrections~\cite{Blumenhagen:2009qh}, can be straightforwardly applied to the singlet $\nu^c$ of the $\mn$.

Thus, one can generate
couplings of the type 
\bea
W_{\text{np}}\sim 
e^{-S^{\text{cl}}} (\nu^c)^3\equiv \kappa\ (\nu^c)^3, 
\label{cubicc3}
\eea
with $\kappa\lsim 1$, if one manages to construct a model with
$n=3$ and
the appropriate suppression factor.
{In fact, the presence of several $\nu^c_i$ superfields in the $\mn$, in contrast to the NMSSM which contains only one singlet superfield $S$ makes easier to generate from this mechanism the parameters $\kappa$ 
(and $\lambda$)
capable to generate a viable mass spectrum.}
Besides, the simultaneous presence of other instanton-induced terms with differing $n$, such as the tadpole term in (\ref{cubicc2}) with $n=1$, can be achieved~\cite{Blumenhagen:2009qh}
by superpotential corrections from multiple instantons with different intersection numbers with the gauge branes.

\section{Conclusions}
\label{conclusion}

We have discussed the domain wall and tadpole problems in the $\mn$, where RH neutrinos are added to the SM spectrum.
Their solutions present peculiarities with respect to the usual ones
found in the NMSSM. An important argument concerns
the presence of several different couplings involving singlet superfields in the $\mn$ (the RH neutrinos), unlike the NMSSM where only one extra singlet is present. In addition, embedding the $\mn$ in the framework of superstring models makes that not all gauge invariant terms are necessarily allowed, providing us with more flexibility for only the interesting terms to be present.

In particular, dimension-4 non-renormalizable terms in the superpotential can be large enough as to solve the domain wall problem through their contributions to the effective potential.
Although they also generate tadpole divergences giving rise to linear terms in the effective potential, these contributions can be small enough for several ranges of the couplings avoiding the destabilization of the hierarchy, as discussed in Sec.~\ref{sec4.1}.

Alternatively,
terms of higher dimensions, odd (even) in the superpotential (the K\"ahler potential), can also be used to solve these problems in the superstring inspired $\mn$
without relying in arguments based on target space duality or $R$-symmetry. 
Unlike dimension-4 terms, their contributions to the effective potential are not large enough to solve the domain wall problem (unless the cut-off scale of the high-energy theory is smaller than the Planck scale).
Nevertheless, the linear contributions generated from tadpole divergences can be large enough to solve the problem without destabilizing the hierarchy, as explained in Sec.~\ref{sec4.2}.

In addition, tadpole terms in the superpotential of the correct order of magnitude to solve the problems can also be generated by instanton effects in type II superstring or F-theory,
as discussed in Sec.~\ref{nonper}.

Finally, 
given the special role of RH neutrinos in the $\mn$ for solving the $\mu$- and $\nu$-problems, simultaneously avoiding the potential domain wall and tadpole problems, 
we have analyzed one of their relevant features in string models. In particular, they are expected to have typically extra $U(1)$ charges at high energies. We have discussed in 
Sec.~\ref{neutrinos}
several mechanisms that still allow the presence of cubic neutrino terms in the low-energy theory, such as 
the mixing between fields due to Fayet-Iliopoulos breaking, the presence of appropriate non-renormalizable terms, or instanton effects. 
The presence of cubic neutrino terms together with Dirac neutrino Yukawas, makes very easy the
generation of a EW-scale seesaw able to reproduce neutrino masses and mixings in agreement with data.  
We thus highly motivate our string model builder colleagues to take account of this option seriously.

\begin{acknowledgments}

We thank \'Angel Uranga for useful conversations about instanton-induced operators in strings.
The work 
of D.L. was supported by the Argentinian CONICET, and also through PIP 11220170100154CO.
The work of C.M. was partially supported by the Spanish 
Research Agency AEI
through grants IFT Centro de Excelencia Severo Ochoa CEX2020-001007-S, PGC2018-095161-B-I00 and PID2021-125331NB-I00; All of them are funded by MCIN/AEI/10.13039/501100011033 and the
second grant also by ERDF "A way of making Europe".

\end{acknowledgments}


\bibliographystyle{utphys}
\bibliography{main.bbl}
\end{document}